\documentclass[11pt]{article}
\usepackage{moriond,epsfig}

\newcommand{\wo}{\omega_0}

\begin{document}
\title{Heisenberg constraints on mesoscopic and molecular
amplifiers}\author{\underline{U. Gavish}$^1$, B. Yurke$^2$, Y.
Imry$^{3}$}

\address{$^1$LKB, Ecole Normale Superieure, Paris }
\address{$^2$Bell Labs, Lucent Technologies, Murray Hill,  NJ}
\address{$^3$Condensed Matter Physics Dept., Weizmann Institute, Rehovot}

\maketitle \abstracts{Heisenberg uncertainty relations for current
components impose constraints on the performance of linear
amplifiers.  Here we derive such constraints for amplifiers in
which the input signal modulates a bias current in order to
produce an amplified output. These amplifiers include transistors,
macroscopic, mesoscopic, or molecular, operated as linear
amplifiers.}

\section{Introduction}

Commutation relations or Heisenberg uncertainty relations for observable
associated with the inputs and outputs of amplifiers have played
an important role in determining the optimum performance that
can be achieved by amplifiers and detectors\cite{shimoda}$^-$\cite{girvin}.
Most of these discussions have
focused on maser, laser, and optical parametric amplifiers, the
first devices to achieve nearly quantum limited performance.
The arguments that establish
the quantum limited performance of amplifiers of the electromagnetic
field, such as optical amplifiers, do not directly carry over to
devices that employ fermionic currents.  It is thus worth addressing
the issue of the quantum limits of amplifier performance in a way
that is directly applicable to semiconductor devices, particularly
since semiconductor device development, such as in the case of single electron transistors
\cite{devoret set,lahoye},
has proceeded to the point where quantum limited
performance seems to be within reach.
Here we present further results in our investigation\cite{Gavish gener constr} of quantum mechanical restrictions on
transistor amplifier performance.

Let $I_{in}$ denote the current delivered by a signal source to the
input of an amplifier and $I_{out}$ denote the current delivered by
the amplifier to a load.  Ideally, the relation between these two
currents would be
\begin{eqnarray} \label{2}
I_{out}(t)=G_p \sqrt{\frac{g_l}{g_s}} I_{in}(t),
\end{eqnarray}
where $G^2_p$ is the power gain and $g_s$ and $g_l$ are the
\emph{differential conductances} of the source and load
respectively.  However, for $G_p \neq 1$, the current-current
commutation relations required by quantum mechanics cannot be
satisfied. This situation is remedied by replacing Eq.~(2) with
\begin{eqnarray} \label{3}
I_{out}(t)=G_p \sqrt{\frac{g_l}{g_s}}I_{in}(t)+I_N(t),
\end{eqnarray}
where $I_N(t)$ is a current operator associated with noise generated
within the amplifier.  As noise, $I_N$ is independent of $I_{in}$
and the two commute: $[I_{in},I_N]=0.$

We consider the case when the current $I(t)$ is investigated
with detectors that respond over only over a frequency window
$\Delta \omega$ around a center frequency $\omega_0$.
Introducing the fourier transform
\begin{eqnarray}\label{4b}
I(t)=\frac{1}{\sqrt{2\pi}}\int_{-\infty}^{\infty}d\omega
I(\omega)e^{-i\omega t},
\end{eqnarray} the current sensed by the detectors is then given by
\begin{eqnarray}\label{4c}
I(t)=\frac{1}{\sqrt{2\pi}}\bar{I}(\omega_0) +H.c
\end{eqnarray}
where we defined the  band-integrated current transform (analogous to the annihilation operator of the harmonic oscillator) by
\begin{eqnarray}\label{4d}\bar{I}(\omega_0)\equiv \int_{\omega_0\pm \frac{1}{2}\Delta \omega}I(\omega)
e^{-i\omega t}d\omega.\end{eqnarray}
For an ideal power amplification (i.e. when one is interested in
transferring maximum power from the system to the amplifier and
from the amplifier to a load resistor) in a stationary (though
nonequilibrium) state it has been shown\cite{comment} in
Ref. 16 that $I_N$  satisfies
\begin{eqnarray} \label{5a}
\langle [\bar{I}_N(\omega_0),\bar{I}_N^{\dag}(\omega_0)]\rangle
=-(G_p^2-1)\frac{\hbar\omega_0}{2}g_{\ell}\Delta \omega
\end{eqnarray} and
\begin{eqnarray}\label{5}
\Delta I_N^2(t)\ge (G_p^2-1)\frac{\hbar\omega_0}{2}g_l\Delta \nu
\end{eqnarray}
provided $\Delta \nu=\Delta \omega/2\pi << \omega_0$, where
$\Delta A \equiv (\langle A^2 \rangle - \langle A
\rangle^2)^{1/2}.$
Eq.\ref{5} is a very general constraint on the minimal noise added
in linear amplification.
The basic assumptions in its derivation are that
the amplifier is linear (i.e. that Eq.\ref{3} holds), that the
total system is in a stationary state, and that the differential
conductance of the amplifier remains constant independently of the
input current signal.  These assumptions make possible the
derivation through the application
of Kubo's fluctuation-dissipation theorem
\cite{kubo0,Kubo1} generalized to
nonequilibrium steady states \cite{Landau Lifshitz kubo}$^-$\cite{D.Cohen}.

We turn now to considering a more specific class of devices,
namely transistor amplifiers (such as, for example, field effect,
single-electron, or molecular transistors). A typical feature of
these devices is that they operate in a nonequilibrium current
carrying state even in the absence of coupling to an input signal.
This nonequilibrium current is accompanied by \emph{shot-noise} -
the nonequilibrium current fluctuations. It is therefore natural
to ask whether the constraint Eq.\ref{5} can be refined to take
the existence of this noise into account. The positive answer to this
question is stated in the next section, and  derived in
the last one.

\section{Main result}
Consider a specific case of a linear amplifier operating in a
stationary state where current is flowing through it even in the
absence of a coupling to any signal. In
this case it is useful to write $I_N$ in Eq.\ref{3}  as a sum of
two currents:
\begin{eqnarray}\label{8} I_N=I_0+I_n\end{eqnarray}
where  $I_0$
is the current of the amplifier before the coupling interaction
between the signal and the amplifier is turned on
and $I_n$ is the change in $I_N$ due to
switching on the coupling. Assume now that the coupling
is proportional to a small dimensionless parameter, $\gamma.$
Since $I_0$ existed before $\gamma$ was switched on, it is of
zeroth order in $\gamma.$ $I_n$ appeared as a result of the
coupling and therefore it is of higher order in $\gamma.$ Also the
power gain $G_p$ is of higher order in $\gamma$ since no coupling
implies no gain. We assume that $I_n$ is of higher order in
$\gamma$ than the gain. Our main result states that the following
inequality must be satisfied:
\begin{eqnarray}\label{9}
\Delta I_0(t)\Delta I_n(t)\ge \frac{1}{4}G_p^2
\hbar\omega_0g_{\ell}\Delta \nu.\end{eqnarray} Eq.\ref{9} has
several nontrivial consequences. For example, it implies that the
"old" shot-noise
 $\Delta I_0^2(t)$ is \emph{necessary} for an ideal operation of the amplifier
since coupling a device with vanishing shot-noise to a signal will
result in the appearance of "new" shot noise  $\Delta I_n^2(t)$
which should diverge in order to maintain  the
inequality in  Eq.\ref{9}.
\section{Derivation of the Heisenberg constraint}
To derive Eq.\ref{9} we make use of Eq.\ref{5a} twice, first in
the presence and then in the absence of the coupling $\gamma:$
 The current noise in these cases is given by
\begin{eqnarray}\label{10} I_N=I_0+I_n ~~~~\gamma \neq 0 \nonumber\\
I_N=I_0 ~~~~~~~~~\gamma=0\end{eqnarray}
 Inserting these into Eq.\ref{5a} yields
\begin{eqnarray} \label{11}
\langle[\bar{I}_0(\omega_0),\bar{I}_0^{\dag}(\omega_0)]\rangle +
\langle[\bar{I}_n(\omega_0),\bar{I}_0^{\dag}(\omega_0)]\rangle
+\langle[\bar{I}_0(\omega_0),\bar{I}_n^{\dag}(\omega_0)]\rangle
+\langle[\bar{I}_n(\omega_0),\bar{I}_n^{\dag}(\omega_0)]\rangle~~~~~~~~~~~~~~~~\nonumber \\
=-(G_p^2-1)\frac{\hbar\omega_0}{2}g_{\ell}\Delta \omega
\end{eqnarray} and
\begin{eqnarray} \label{12}
\langle[\bar{I}_0(\omega_0),\bar{I}_0^{\dag}(\omega_0)]\rangle
=\frac{\hbar\omega_0}{2}g_{\ell}\Delta \omega
\end{eqnarray}
were we have used the fact that $G=0$ when $\gamma=0.$
Subtracting the last equation from the previous one, one gets
\begin{eqnarray} \label{13}
\langle [\bar{I}_n(\omega_0),\bar{I}_0^{\dag}(\omega_0)]\rangle
+\langle [\bar{I}_0(\omega_0),\bar{I}_n^{\dag}(\omega_0)]\rangle
+\langle [\bar{I}_n(\omega_0),\bar{I}_n^{\dag}(\omega_0)]\rangle
=-G_p^2\frac{\hbar\omega_0}{2}g_{\ell}\Delta \omega.
\end{eqnarray}
Since we assumed that $I_n$ is of higher order in $\gamma$ than
the gain, the term $\langle
[\bar{I}_n(\omega_0),\bar{I}_n^{\dag}(\omega_0)]\rangle$ is of
higher order in $\gamma$ than the three other terms in Eq.\ref{13}.
Since this equation should hold for any value of $\gamma$ small
enough for the amplifier to be regarded as linear, $\langle
[\bar{I}_n(\omega_0),\bar{I}_n^{\dag}(\omega_0)]\rangle$ must
vanish. Thus, Eq.\ref{13} becomes:
\begin{eqnarray} \label{14}
\langle [\bar{I}_n(\omega_0),\bar{I}_0^{\dag}(\omega_0)]\rangle
+\langle [\bar{I}_0(\omega_0),\bar{I}_n^{\dag}(\omega_0)]\rangle
=-G_p^2\frac{\hbar\omega_0}{2}g_{\ell}\Delta \omega.
\end{eqnarray}
We now use the fact that for any pair of hermitian Heisenberg
operators $A_1(t)$ and $A_2(t),$ one has
\begin{eqnarray} \label{15a}
\langle \bar{A}_i(\wo)\rangle=0~~~~~~ ~~\omega_0\neq
0~~~~~~~~~~i=1,2,
\end{eqnarray} and
\begin{eqnarray} \label{15}
\langle \bar{A}_2(\wo)\bar{A}_1(\wo)\rangle=\langle
\bar{A}_1(\wo)\bar{A}_2(\wo)\rangle=0
\end{eqnarray} where
$\bar{A}_i(\wo)=\int_{\omega_0\pm \frac{1}{2}\Delta \omega}d\omega \frac{1}{\sqrt{2\pi}} \int_{-\infty}^{\infty} dt_i e^{i\omega t} A_i(t_i),$ $i=1,2,$
provided that the averages are performed
in a stationary state (the proof of Eq.\ref{15} is straightforward
by substitution of the definition of $\bar{A}_i(\wo),$
 making a change of the integration variables $\tau_1=t_1- t_2, \tau_2=\frac{1}{2}(t_1+ t_2) $ and integrating over $\tau_2$).
Taking $A_1=I_0$ and $A_2=I_N,$ Eq.\ref{15} enables us to rewrite
Eq.\ref{14} in the form of an expectation value of a commutator of
two hermitian operators $\bar{I}_n(\omega_0)+\bar{I}_n^{\dag}(\omega_0)$ and $i(\bar{I}_0^{\dag}(\omega_0)-\bar{I}_0(\omega_0))$ which are analogous
to a position and a momentum operator, respectively, or to the field
quadrature components of quantum optics:
\begin{eqnarray} \label{16}
\langle
[\bar{I}_n(\omega_0)+\bar{I}_n^{\dag}(\omega_0),i(\bar{I}_0^{\dag}(\omega_0)-\bar{I}_0(\omega_0))]\rangle
=-iG_p^2\frac{\hbar\omega_0}{2}g_{\ell}\Delta \omega~.
\end{eqnarray} This implies the uncertainty relation
\begin{eqnarray} \label{17}
\Delta (\bar{I}_n(\omega_0)+\bar{I}_n^{\dag}(\omega_0))\Delta
(i(\bar{I}_0^{\dag}(\omega_0)-\bar{I}_0(\omega_0))) \geq
\frac{1}{2} G_p^2\frac{\hbar\omega_0}{2}g_{\ell}\Delta \omega.
\end{eqnarray} Eqs. \ref{15a} and \ref{15} also imply (together with Eq.\ref{4c}):
\begin{eqnarray} \label{18}
\Delta (\bar{I}_n(\omega_0)+\bar{I}_n^{\dag}(\omega_0))^2= 2\pi \langle  I_n^2(t)\rangle\nonumber\\
\Delta (i(\bar{I}_0^{\dag}(\omega_0)-\bar{I}_0(\omega_0)))^2= 2
\pi \langle  I_0^2(t)\rangle.
\end{eqnarray} Finally, substituting the last two equalities into Eq.\ref{17} one recovers the constraint, Eq.\ref{9}.

To conclude, a novel Heisenberg constraint on shot-noise carrying
linear amplifier was obtained. This constraint relates the device
shot noise before coupling to the signal and the one added due to
this coupling. One consequence of this relation is that an attempt
to indefinitely reduce the shot-noise in the device in the absence
of a signal  will result in the appearance of diverging new
shot-noise after the coupling to the signal is switched on.

\section*{Acknowledgments}
The research at WIS was supported by a Center of Excellence of the
Israel Science Foundation (ISF) and by the German Federal Ministry
of Education and Research (BMBF), within the framework of the
German Israeli Project Cooperation (DIP).


\section*{References}
\begin{thebibliography}{99}
\bibitem{shimoda}
K. Shimoda, H. Takahasi, and C. H. Towns, J. Phys. Soc. Japan
\textbf{12}, 686 (1957).

\bibitem{heffner}
H. Heffner and G. Wade, J. Appl. Phys. \textbf{29}, 1262 (1958).

\bibitem{louisell}
W. H. Louisell, A. Yariv, and A. E. Siegman, Phy. Rev.
\textbf{124}, 1646 (1961).

\bibitem{house}
H. A. Haus and S. A. Mullen, Phys. Rev. \textbf{128}, 2407 (1962).

\bibitem{takahasi}
H. Takahasi, {\it Advances in Communication Systems}, ed. A. V.
Balakrishnan (Academic, New York, 1965).

\bibitem{caves1} C. M. Caves, K. S. Thorne, R. Drever, V.  Sandberg, and M. Zimmermann,
 Rev. Mod. Phys. \textbf{ 52}, 341 (1980).

\bibitem{caves2} C. M. Caves, Phys. Rev. D \textbf{26}, 1817 (1982).

\bibitem{devoret set} See e.g. M. H. Devoret and R. J. Schoelkopf, Nature \textbf{406}, 1039 (2000) and references therein.

\bibitem{Yurke denker}
B. Yurke and J. S. Denker, Phys. Rev. \textbf{A 29}, 1419 (1984).

\bibitem{yuen}
H. P. Yuen, Phys. Rev. A \textbf{13}, 2226 (1976).

\bibitem{hollenhorst}
J. N. Hollenhorst, Phys. Rev. D \textbf{19}, 1669 (1979).


\bibitem{lahoye}
M. D. LaHoye, O. Buu, B. Camarota, and K. C. Schwab,
Science \textbf{304}, 74 (2004).


\bibitem{averin} D. V. Averin, cond-mat/0301524.
\bibitem{girvin} A. A. Clerk, S. M. Girvin and A. D. Stone Phys. Rev. B \textbf{67}, 165324 (2003).


\bibitem{comment} Eqs.\ref{5a} and \ref{5} here are immediately obtained from Eqs.11,12, 26-30, 35 and 37 in Ref. 16.


\bibitem{Gavish gener constr} U. Gavish, B. Yurke and Y. Imry
"Generalized Constraints on Quantum Amplification" cond-mat/0407415.

\bibitem{kubo0} R. Kubo, Can. J. Phys. 34, 1274 (1956).

\bibitem{Kubo1} R. Kubo, J. Phys. Soc. Japan, \textbf{12}, 570 (1957);

\bibitem{Landau Lifshitz kubo}
 L. D. Landau and E. M. Lifshitz, \textit{Statistical Physics Part 1}, 3rd ed.,
Sec. 126, Butterworth Heinemann (1997).

\bibitem{moriond 2001} U. Gavish, Y. Levinson
and Y. Imry, Proc. of  Recontres de Moriond 2001: Electronic
Correlations: from Meso- to Nanophysics, T. Martin et al., eds.
EDPScience 2001.

\bibitem{Canary} U. Gavish, Y. Imry, and B. Yurke,   Vol. 5449, 257, Proc. SPIE Noise Conference, Grand
Canary, 2004, cond-mat/0404270.

\bibitem{D.Cohen} D. Cohen, Ann. Phys. \textbf{283}, 175 (2000).


\end{thebibliography}
\end{document}